\documentclass[journal,twocolumn,letterpaper]{IEEEJERM}

\ifCLASSINFOpdf
\else
\fi

\usepackage{times,amsmath,epsfig}
\usepackage{fancyhdr}
\usepackage{amsmath}
\usepackage{amsfonts}
\usepackage{amssymb}
\usepackage{array}
\usepackage{graphicx}
\usepackage{url}
\usepackage{bm}
\usepackage{breqn}
\usepackage{xcolor}
\usepackage{soul}
\usepackage{amssymb}
\usepackage[breaklinks=true]{hyperref}
\usepackage{breakcites}

\usepackage{gensymb}
\usepackage{threeparttable}
\usepackage{multicol}
\usepackage{mathrsfs}
\usepackage{multirow}
\usepackage{array}
\usepackage{booktabs}
\usepackage{adjustbox}
\usepackage{subcaption} 
\usepackage[final]{changes}

\begin{document}

%

\vspace{1 cm}

\title{Deep Learning-Based Automatic Diagnosis System for Developmental Dysplasia of the Hip}

%
\author{Yang~Li$^{1,2*}$, 
        Leo~Yan~Li-Han$^{3*}$,
        Hua~Tian$^{1,2}$, 
       
}

\markboth{IEEE Journal of Translational Engineering in Health and Medicine}
{Y. Li \MakeLowercase{\textit{et al.}}: Deep Learning-Based Automatic Diagnosis System for Developmental Dysplasia of the Hip}

\twocolumn[
\begin{@twocolumnfalse}
  
\maketitle


\begin{abstract}
 
Objective: \textnormal{The clinical diagnosis of developmental dysplasia of the hip (DDH) typically involves manually measuring key radiological angles---Center-Edge (CE), T\"{o}nnis, and Sharp angles---from pelvic radiographs, a process that is time-consuming and susceptible to variability. This study aims to develop an automated system that integrates these measurements to enhance the accuracy and consistency of DDH diagnosis.} Methods and procedures: \textnormal{We developed an end-to-end deep learning model for keypoint detection that accurately identifies eight anatomical keypoints from pelvic radiographs, enabling the automated calculation of CE, T\"{o}nnis, and Sharp angles. To support the diagnostic decision, we introduced a novel data-driven scoring system that combines the information from all three angles into a comprehensive and explainable diagnostic output.} Results: \textnormal{The system demonstrated superior consistency in angle measurements compared to a cohort of eight moderately experienced orthopedists. The intraclass correlation coefficients for the CE, T\"{o}nnis, and Sharp angles were 0.957 (95\% CI: 0.952--0.962), 0.942 (95\% CI: 0.937--0.947), and 0.966 (95\% CI: 0.964--0.968), respectively. The system achieved a diagnostic F1 score of 0.863 (95\% CI: 0.851--0.876), significantly outperforming the orthopedist group (0.777, 95\% CI: 0.737--0.817, $p = 0.005$), as well as using clinical diagnostic criteria for each angle individually ($p<0.001$).} Conclusion: \textnormal{The proposed system provides reliable and consistent automated measurements of radiological angles and an explainable diagnostic output for DDH, outperforming moderately experienced clinicians.} 

Clinical impact: \textnormal{This AI-powered solution reduces the variability and potential errors of manual measurements, offering clinicians a more consistent and interpretable tool for DDH diagnosis.}  \end{abstract}


\end{@twocolumnfalse}]

{
  \renewcommand{\thefootnote}{}%
  \footnotetext[1]{$^*$Yang Li and Leo Yan Li-Han contributed equally to this work and designated as co-first authors.}
  \footnotetext[2]{$^1$Department of Orthopedics, Peking University Third Hospital, Beijing, China, 100191.}
  \footnotetext[3]{$^2$Engineering Research Center of Bone and Joint Precision Medicine, Ministry of Education, Beijing, China, 100191.}
  \footnotetext[4]{$^3$The Edward S. Rogers Sr. Department of Electrical \& Computer Engineering, University of Toronto, 10 King's College Rd, Toronto, ON M5S 3G4, Canada.}
  \footnotetext{Corresponding author: Hua Tian (tianhua@bjmu.edu.cn)}
}
 
%
\IEEEpeerreviewmaketitle

\section{Introduction}

\IEEEPARstart{D}{evelopmental} dysplasia of the hip (DDH) is a group of hip disorders primarily characterized by a shallow acetabulum and inadequate coverage of the femoral head. The global prevalence of DDH varies between $0.15\%$ to $3.5\%$, depending on the diagnostic methods and criteria~\cite{siegel2011pediatric, dezateux2007developmental, tian2017prevalence, tao2023prevalence}. DDH is one of the leading causes of osteoarthritis~\cite{sharp1961acetabular} and accounts for up to $29\%$ of hip arthroplasty performed in \replaced{adult patients}{individuals} younger than 60 years~\cite{dezateux2007developmental}. While common symptoms include pain and limping, mild cases of DDH may remain asymptomatic, leading to delayed or missed diagnosis~\cite{dezateux2007developmental}. Such delays can further complicate treatment and increase the risk of failure~\cite{sewell2009developmental}, underscoring the importance of timely and accurate diagnosis to preserve patient quality of life.

Radiography is the cornerstone imaging modality of DDH diagnosis. Based on radiographic assessments, appropriate therapeutic strategies or interventional procedures can be determined for different stages of the disease~\cite{pereira2014recognition}. As such, several radiological indices have been developed to assist in diagnosing DDH from pelvic radiographs. Among these, the Center–Edge (CE) angle of Wiberg assesses the lateral coverage of the acetabulum, with a CE angle of less than 20\degree considered indicative of DDH~\cite{wiberg1939studies}. The T\"{o}nnis angle, also known as the acetabular index, evaluates the weight-bearing surface of the acetabulum, with a normal range from 0\degree to 10\degree~\cite{tonnis1976normal}. Additionally, the Sharp angle (or acetabular angle) describes the inclination of the acetabulum, with an angle greater than 47\degree suggesting the presence of DDH~\cite{sharp1961acetabular}. 

\begin{figure*}[!t]
	\centering
	\includegraphics[width=0.65\linewidth]{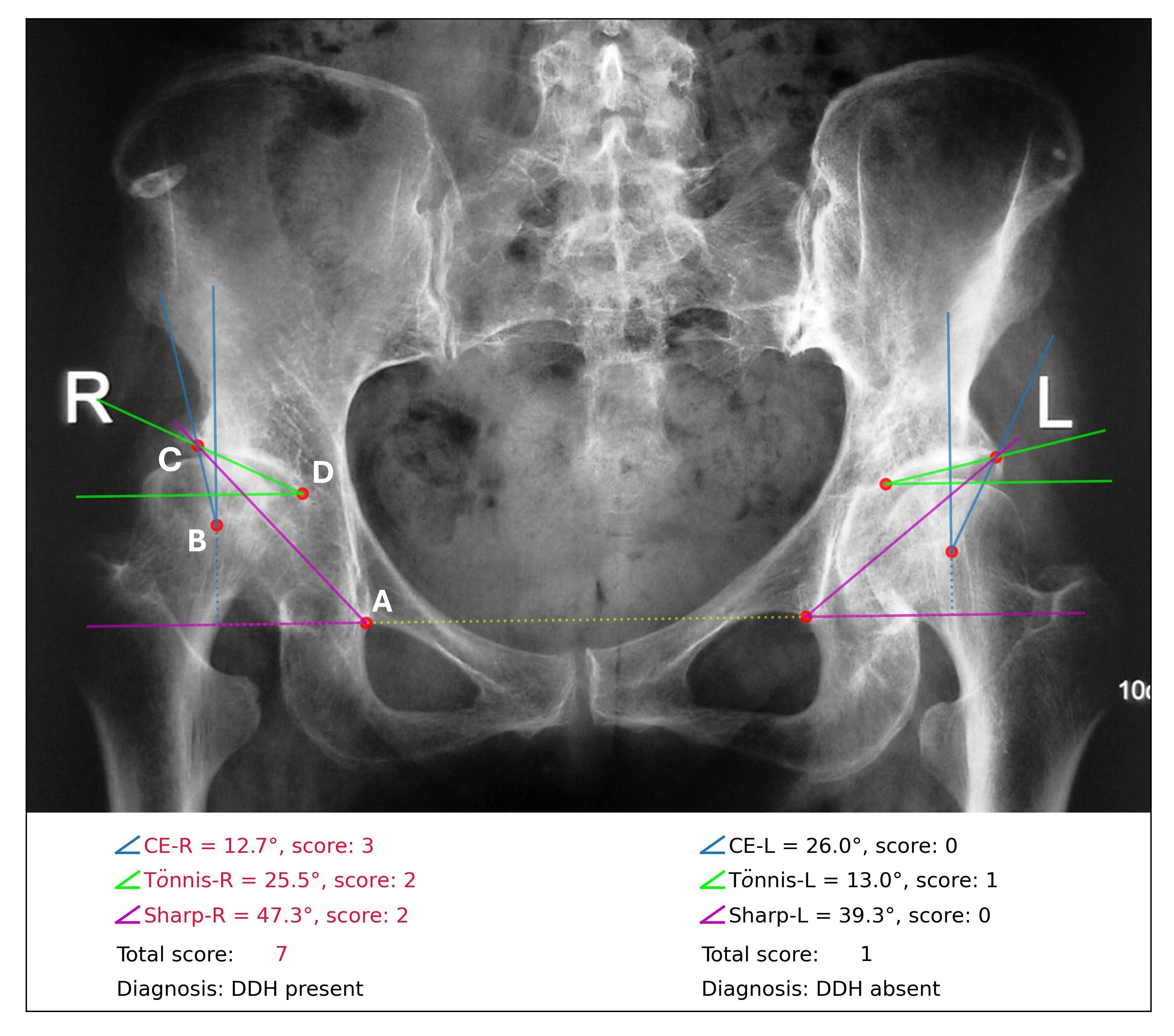}
	\caption{Diagnosis generated by the proposed system based on an anteroposterior view pelvic radiograph. The system detects four keypoints on each side of the hip: (A) the inferior boundary of the teardrop point, (B) center of the femoral head, (C) lateral edge of the acetabulum, and (D) medial aspect of the acetabulum. The angle measurements and diagnostic scores are displayed in the bottom text (CE: Center–Edge). Angles that exceed the normal range are highlighted in red in the textual results. The right hip (marked as R on the radiograph) is diagnosed as ``DDH present”, as the total score (7) is greater than the diagnostic threshold of 5. The diagnosis for the left hip (marked as L on the radiograph) is ``DDH absent”.}
	\label{fig:1}
\end{figure*}

However, the accurate measurement of these diagnostic indices depends on the manual identification and assessment of key landmarks in radiographs, a process that can be inefficient and prone to errors, especially for less experienced clinicians. Consequently, diagnostic accuracy is often compromised by measurement variability and the quality of the radiographs~\cite{nelitz1999reliability}. Moreover, the subtle morphological differences between mild DDH and normal hips or other conditions can further complicate the diagnosis (see the minor difference between left and right hip shown in Figure~\ref{fig:1}), necessitating extensive training and clinical experience. To enhance diagnostic sensitivity, clinicians are suggested to comprehensively interpret the CE, T\"onnis, and Sharp angles before making a diagnosis~\cite{pereira2014recognition}\added{\cite{welton2023radiographic}}, as these indices provide complementary insights into the condition. However, there is a lack of standardized and objective clinical guidelines for integrating those measurements into a definitive DDH diagnosis, highlighting the need for a reliable, interpretable, and automated diagnostic approach.

Deep learning algorithms have shown considerable promise in analyzing pelvic radiographs across various applications, including fracture detection~\cite{badgeley2019deep}, osteonecrosis diagnosis and staging~\cite{li2020deep}, and radiological feature measurement~\cite{rouzrokh2021deep}. In the context of DDH diagnosis, Park~\textit{et al.}~\cite{park2021diagnostic}, \added{Den~\textit{et al.}~\cite{den2023diagnostic}, and Magn{\'e}li~\textit{et al.}~\cite{magneli2024application}} developed \deleted{a} convolutional neural network\added{s} (CNN) to \added{respectively} detect \deleted{pediatric} DDH from \added{pediatric and adult} pelvic radiographs, achieving performance comparable to that of \replaced{clinicians}{radiologists}. However, the CNN model\added{s} operated as \deleted{a} ``black box" \added{classifiers}, lacking the clinical interpretability essential for decision-making. Li~\textit{et al.}~\cite{li2019auxiliary} used a modified Mask-RCNN model~\cite{he2017mask} to identify 4 keypoints on pelvic radiographs, from which the Sharp angle was calculated to diagnose DDH. Although the model achieved diagnostic accuracy comparable to that of surgeons, relying on a single index may not provide a comprehensive assessment. Therefore, it is important to combine multiple indices for a more reliable diagnosis~\cite{pereira2014recognition}\added{\cite{welton2023radiographic}}. In another study, Yang~\textit{et al.}~\cite{yang2020feasibility} proposed a CNN model with hourglass architecture to predict probability maps for 10 keypoints on pelvic radiographs\deleted{, which allows for the calculation of CE, T\"onnis, and Sharp angles}. \added{Similarly, Li~\textit{et al.}~\cite{li2024deep} developed a Vnet-based~\cite{milletari2016v} model that automatically recognize 4 keypoints from each side of the hip. Both approaches allowed for an automatic calculation of CE, T\"onnis, and Sharp angles.} While \replaced{those}{this} model\added{s} demonstrated \replaced{promising}{state-of-the-art} performance in keypoint detection \added{and angle measurements}, \replaced{they}{it} did not integrate the measurements into a unified diagnostic outcome, which may limit \replaced{their}{its} clinical utility. 

In this study, we propose an end-to-end system for the comprehensive diagnosis of \added{adult} DDH using anteroposterior view pelvic radiographs. Specifically, we developed a keypoint detection model based on the Mask-RCNN architecture to detect 8 keypoints on each pelvic radiograph. Subsequently, the CE, T\"onnis, and Sharp angles are automatically measured according to the detected keypoints and their clinical definitions. To provide a more robust diagnosis, we introduced a new data-driven scoring system that integrates these angle measurements for a comprehensive assessment of DDH. Figure~\ref{fig:1} illustrates an example of the visualized results generated by our system, showing a diagnosis of ``DDH present" in the right hip and ``DDH absent" in the left hip.

The remainder of this paper is organized as follows: Section II details the methods used in this study, including data collection, the keypoint detection model, the data-driven scoring system, as well as evaluation metrics. Section III presents the experimental results. Section IV discusses the findings, and Section V concludes the study.

\section{Methodology}

\subsection*{Data}

This study used a retrospective set of anteroposterior view pelvic radiographs sourced from the radiology repository of Peking University Third Hospital. We reviewed radiographs from patients over 18 years old who presented with developmental dysplasia of the hip (DDH) at the orthopedic clinic between 2020 and 2022. Pediatric radiographs \added{were excluded due to distinct radiological characteristics and clinical management strategies.} \replaced{Moreover,}{as well as} radiographs exhibiting fractures, internal fixation, prostheses, \added{or conditions affecting radiological measurements of the hip} were excluded from the analysis. Additionally, cases with severe osteoarthritis and advanced osteonecrosis (stage III and IV)\deleted{, or any condition preventing the identification of the femoral head center} were also excluded\added{, as these conditions cause significant anatomical alterations, making radiological measurements less clinically relevant}. After applying these criteria, 1,683 pelvic radiographs, corresponding to 3,366 hips, were included in the study. Of these, 150 radiographs (300 hips) were reserved \added{exclusively} for testing (denoted as the Test set), while the remaining 1,533 radiographs (3,066 hips) were used for \added{model} training\replaced{,}{and} validation\added{, and hyperparameter tuning} (denoted as the \replaced{Train-Val}{training} set). This study was conducted adhering to the tenets of the Declaration of Helsinki.

Data annotation was conducted by three orthopedic surgeons, each with at least 15 years of clinical and surgical experience. Using a locally hosted open-source annotation tool~\cite{makesense2019}, the annotators labeled four keypoints on each hip (eight per radiograph), as shown in Figure~\ref{fig:1}: (A) the inferior boundary of the teardrop point, (B) the center of the femoral head, (C) the lateral edge of the acetabulum, and (D) the medial aspect of the acetabulum. In addition to the keypoints, a bounding box containing the entire pelvic region, which included all eight keypoints, was also marked. This bounding box was used to guide the model in focusing on the region of interest during training. 

Each surgeon independently annotated the radiographs, and the coordinates of each labeled point and bounding box were averaged across the three annotators to establish the ground truth. To estimate measurement variability, all annotators repeatedly labeled radiographs of the Test set five times (with a 2-day interval). These repeated measurements were then used in performance evaluation, representing the expected variability among human experts. Lastly, each annotator provided a binary diagnosis for each hip (i.e., ``DDH present” or ``DDH absent”) based on their measurements and clinical assessments. In cases of diagnostic disagreement, a majority vote determined the final diagnosis. 

Following established clinical guidelines~\cite{sharp1961acetabular, pereira2014recognition, wiberg1939studies, tonnis1976normal, welton2023radiographic, hanson2015discrepancies}, the radiological measurements in this study were defined as follows, referring to Figure~\ref{fig:1}. The \textbf{Horizontal reference line} (yellow dotted line) was defined as the line connecting the two teardrop points and passing through point A. The \textbf{Vertical reference line} (blue dotted line) was the vertical line perpendicular to the horizontal reference line and passing through point B. The \textbf{Center-Edge (CE) angle} was defined as the angle (blue) between the line connecting points B and C and the vertical reference line. The \textbf{T\"onnis angle} was defined as the angle (green) between the line connecting points C and D and the line parallel to the horizontal reference line and passing through point D. Finally, the \textbf{Sharp angle} was defined as the angle (purple) between the line connecting points A and C and the horizontal reference line. The ground truth measurement of CE, T\"onnis, and Sharp angles were calculated based on the ground truth keypoint locations and these defined measurement criteria.

\subsection*{Keypoint Detection}

We developed a keypoint detection model based on the Mask-RCNN architecture~\cite{he2017mask}, with a Resnet-50 network~\cite{he2016deep} as the feature extraction backbone. Input radiographs were passed through the ResNet-50 network, producing feature maps that were subsequently fed into the region proposal network to generate candidate regions of interests (RoI) corresponding to the pelvic area. The proposed RoIs were refined using the RoIAlign module, which converts them into fixed-size feature maps. Then, two parallel branches processed these aligned features for keypoint detection and bounding box regression, respectively. Unlike the original Mask-RCNN model designed for object segmentation, we redefined the output to detect keypoints by creating ``one-hot" masks, where only one pixel at the keypoint location has a value of 1, and all other pixels are set to 0. Additionally, the object classification branch in the original Mask-RCNN model was removed, as our task only involves a single class (i.e., the pelvis region). On the other hand, the bounding box regression branch was retained to facilitate RoI identification, thereby improving the keypoint detection performance. 

The loss function used for training the model was defined as the sum of the keypoint detection loss ($L_{kp}$) and bounding box regression loss ($L_{box}$), such that $L=L_{kp} + L_{box}$. Given that only one foreground pixel corresponds to each keypoint, we employed focal loss~\cite{lin2017focal} as the keypoint detection loss instead of the binary cross-entropy loss, as it improves both training efficiency and accuracy by focusing the model on harder-to-classify examples. Focal loss modulates the cross-entropy loss by down-weighting easily classified samples and emphasizing difficult cases. The keypoint detection loss in our model was defined as:
\begin{equation}
	L_{kp} =-\frac{1}{N}\sum_{k=1}^K\sum_{i=1}^H\sum_{j=1}^W \begin{cases} 
		(1-p_{kij})^{\gamma}\log{p_{kij}} & \text{if }y_{kij}=1, \\
		p_{kij}^\gamma\log{(1-p_{kij})} & \text{Otherwise}
	\end{cases}
\end{equation}
where $p_{kij}$ is the model's predicted probability that pixel ($i,j$) belongs to keypoint $k$, $y_{kij}$ is the ground truth label for pixel ($i,j$), $\gamma$ is the focusing parameter (set to 2, as per~\cite{lin2017focal}), $K$ is the total number of keypoints, and $H$ and $W$ are the height and width of the image, respectively.

For the bounding box regression, we employed the Smooth L1 loss, which is more robust to outliers than the L2 loss (i.e., the Mean Square Error). The bounding box regression loss was defined as:
\begin{equation}
	L_{box} = \sum_{i\in {x,y,w,h}} \text{Smooth L1}(t_i - \hat{t}_i)
\end{equation}
where $t_i$ and $\hat{t}_i$ represent the ground truth and predicted bounding parameters, specifically the coordinates $(x, y)$ of the top left corner, width ($w$), and height ($h$) of the box.

\begin{figure*}[t]
	\centering
	\includegraphics[width=0.86\linewidth]{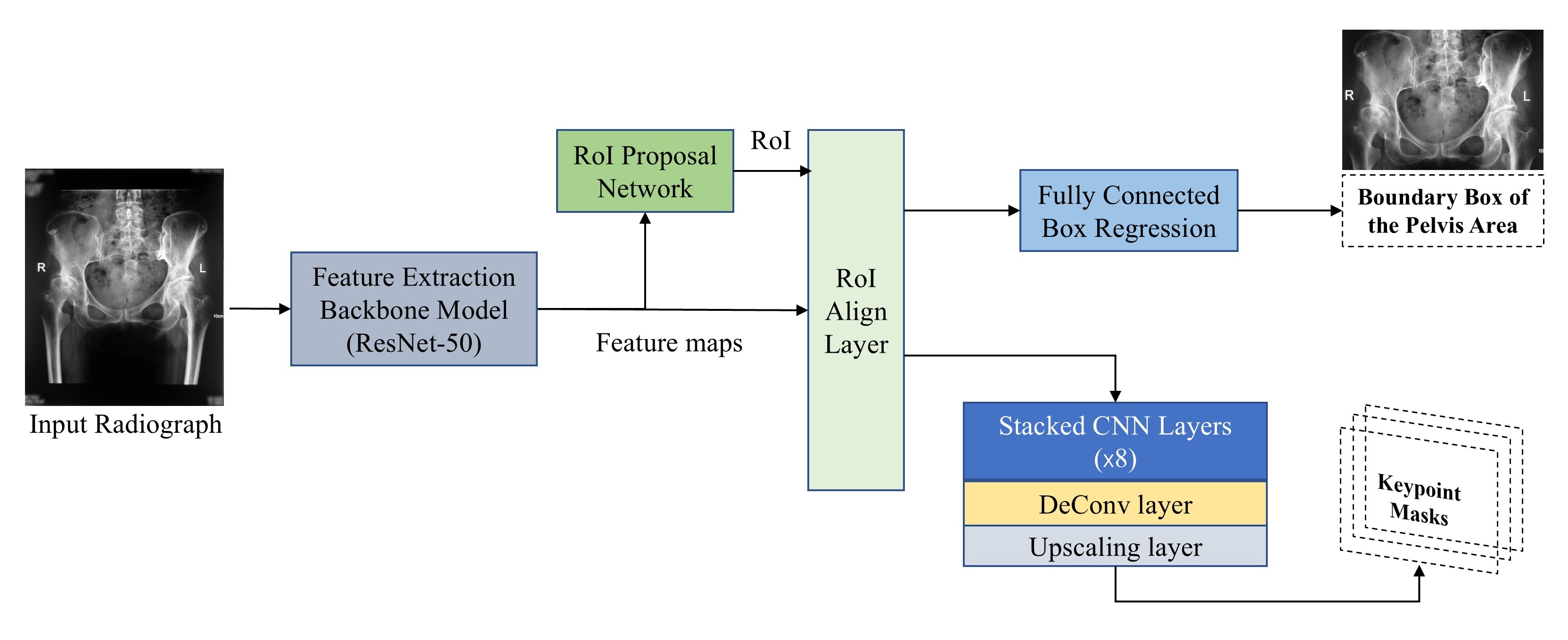}
	\caption{The architecture of the keypoint detection model. The ResNet50 model was used to extract features from the input radiograph. The feature maps were then fed into the region proposal network to generate candidate regions of interest (RoI). The RoIAlign layer converts the feature maps and proposed regions of interest into the same size. Subsequently, two parallel neural network branches are responsible for keypoint detection and bounding box regression, respectively.}
	\label{fig:2}
\end{figure*}

Figure~\ref{fig:2} provides an overview of the keypoint detection model architecture. During training, we used an initial learning rate of 0.005, which was reduced by a factor of 5 when the validation loss plateaued for three consecutive epochs. The model was trained for 15 epochs with a mini-batch size of 4, using the stochastic gradient descent optimizer with a weight decay of 0.0001 and momentum of 0.9. Standard data augmentation techniques, such as small-angle rotation and adding random noise, were used to increase data diversity and model generalizability. To examine robustness, we applied 10-fold cross-validation (CV) for performance evaluation. In the inference phase, the models trained on each CV fold were tested on the Test set, and the performance confidence interval (CI) was recorded and reported.

\begin{table}[!t]
	\caption{The Diagnosis Scoring System for Developmental Dysplasia of the Hip}
	\label{table:1}
	\centering
	\resizebox{0.99\columnwidth}{!}{
		\begin{threeparttable}
			\begin{tabular}{@{}lllll@{}}
				\toprule
				\textbf{Classes}    & \textbf{CE angle}      & \textbf{T\"{o}nnis angle} & \textbf{Sharp angle} & \textbf{Score} \\ \midrule
				Normal     &\textgreater $25\degree$   &\textless $10\degree$ &\textless $42\degree$ &0  \\
				Borderline &$20\degree$--$25\degree$   &$10\degree$--$13\degree$  &$42\degree$--$47\degree$  &1        \\
				DDH        &\textless $20\degree$  &\textgreater $13\degree$  &\textgreater $47\degree$    &\begin{tabular}[c]{@{}l@{}}3 for CE angle\\ 2 for others\end{tabular}        \\ \bottomrule
			\end{tabular}
			\begin{tablenotes}
				\item Note: the diagnosis is ``DDH present" when the total score from three angles is $\geq5$; otherwise, the diagnosis is ``DDH absent".
			\end{tablenotes}
		\end{threeparttable}
	}
\end{table}

\subsection*{Scoring System for DDH Diagnosis}

Previous studies suggest that combining the CE, T\"{o}nnis, and Sharp angles provides a more sensitive diagnostic approach, particularly for mild cases of developmental dysplasia of the hip (DDH)~\cite{pereira2014recognition}\added{\cite{welton2023radiographic}}. However, to the best of our knowledge, no formal clinical guidelines currently exist for integrating these measurements. To address this gap, we developed a new data-driven scoring system that offers a quantitative and objective diagnosis of DDH by incorporating these three angular measurements. 

In our scoring system, each hip is categorized into one of three classes (i.e., normal, borderline, and DDH) based on the clinical guideline for each of the three angles. As such, each hip receives three diagnoses, one from each angle. The classification criteria for each angle are as follows:

\begin{itemize}
	\item \textbf{CE angle}: Normal is defined as \textgreater $25\degree$, borderline as $20\degree-25\degree$, and DDH as \textless $20\degree$~\cite{wiberg1939studies}.
	
	\item \textbf{T\"{o}nnis angle}: Normal is \textless $10\degree$, borderline is $10\degree-13\degree$, and DDH is \textgreater $13\degree$~\cite{pereira2014recognition}\cite{fa2014superiority}.
	
	\item \textbf{Sharp angle}: Normal is \textless $42\degree$, borderline is $42\degree-47\degree$, and DDH is \textgreater $47\degree$~\cite{sharp1961acetabular}.
\end{itemize}

Then, each diagnosis from the three angles is assigned a corresponding score/weight. Specifically, a score of 0 is given for normal classifications across all angles, and a score of 1 is assigned for borderline cases. For DDH diagnoses, the CE angle receives a score of 3, while the T\"{o}nnis and Sharp angles are assigned scores of 2. The total score from the three angles is then summed, and the final diagnosis is made based on a decision threshold. If the total score is $\geq 5$, the hip is diagnosed as ``DDH present"; otherwise, the diagnosis is ``DDH absent."

To determine the optimal parameters of the scoring system, we performed grid search in the \replaced{Train-Val}{training} set to fine-tune the scores assigned to each angle and the diagnostic decision threshold. \added{Like the technique used in keypoint detection, the parameter search was performed in a 10-fold cross-validation manner to prevent potential overfitting and enhance the robustness of selected diagnostic parameters.} \replaced{For each CV fold, the}{The} optimization aimed to maximize diagnostic performance between the scoring system and the ground truth labels. In this study, the diagnostic performance was quantified using the $\text{F1-score} =\frac{2\text{TP}}{2\text{TP}+\text{FP}+\text{FN}}$, a single-value metric robust to imbalanced data distribution. Then, the final\deleted{ized} criteria \added{were determined by selecting the thresholds that maximized the average diagnostic performance across the 10-fold cross-validation while minimizing the variance.} \replaced{The detailed}{and} parameters of the scoring system are summarized in Table~\ref{table:1}.

\subsection*{Performance Evaluation}

The keypoint detection performance was evaluated using the object keypoint similarity (OKS) metric~\cite{ruggero2017benchmarking}, which measures the normalized distance between predicted and ground-truth keypoints. An OKS score of 1 indicates a perfect keypoint detection, while scores closer to 0 reflect increasing deviation from the ground-truth location. Following the convention in~\cite{lin2014microsoft}, detection precision and recall were assessed by thresholding OKS scores. Specifically, a keypoint prediction was considered a true positive if the OKS value exceeded a specified threshold; otherwise, it was deemed a false negative. By further varying the OKS threshold from 0.5 to 0.95 in steps of 0.05, we calculated the mean average precision (mAP) and mean average recall (mAR) as metrics for keypoint detection. 

Additionally, sensitivity analyses were performed to evaluate the influence of various model design choices on the keypoint detection performance. We compared the detection mAP and mAR using different loss functions (focal loss vs. cross-entropy loss), feature extraction backbone models (ResNet vs. ResNeXt~\cite{xie2017aggregated} vs. Feature Pyramid Network~\cite{lin2017feature}), and types of keypoint masks (binary mask vs. heatmap mask~\cite{yang2020feasibility}\cite{tompson2015efficient}\cite{law2018cornernet}) to determine the optimal model configuration.

To evaluate the accuracy of the angle measurements, Bland-Altman analysis was employed to quantify the agreement between angles calculated from the predicted and the ground-truth keypoints. To further benchmark the model's performance against human experts, we recruited another group of eight orthopedic clinicians who did not participate in data annotation and had \replaced{moderate}{more than six years of} clinical and surgical experience \added{(six to ten years)} to manually mark the keypoints and diagnose the radiographs in the Test set. Subsequently, the intraclass correlation coefficients (ICC)~\cite{koo2016guideline} were computed to compare the consistency between ground-truth angle measurements and those generated by our model, the original annotators (from repeated annotations), the orthopedists, and state-of-the-art results from \deleted{a} previous stud\replaced{ies}{y}~\cite{yang2020feasibility, li2024deep}.

Lastly, the performance of the DDH diagnosis was assessed by comparing the F1 score of the proposed scoring system with those of the clinician groups, as well as with the diagnostic criteria based on individual angular measurements. The Mann-Whitney U test was employed to analyze the statistical significance of the comparisons.

\begin{table}[!t]
	\caption{Data Characteristics}
	\label{table:3}
	\centering
	\resizebox{0.99\columnwidth}{!}{
		\begin{threeparttable}
			\begin{tabular}{@{}lllll@{}}
				\toprule
				\multirow{2}{*}{} & \multirow{2}{*}{\textbf{\begin{tabular}[c]{@{}l@{}}Radiograph\\ count\end{tabular}}} & \multirow{2}{*}{\textbf{\begin{tabular}[c]{@{}l@{}}Hip\\ count\end{tabular}}} & \multicolumn{2}{l}{\textbf{Hip diagnosis (count [percentage])}} \\ \cmidrule(l){4-5}   & & & 
				\textbf{DDH$^a$ absent} & \textbf{DDH present} \\ \midrule
				All         & 1683  & 3366  & 3024 (89.8\%)   & 342 (10.2\%) \\
				Train-Val   & 1533  & 3066  & 2758 (90.0\%)   & 308 (10.0\%) \\
				Test        & 150   & 300   & 266  (88.7\%)   & 34  (11.3\%)  \\
				\bottomrule
			\end{tabular}
			\begin{tablenotes}
				\item $^a$ DDH denotes developmental dysplasia of the hip. 
			\end{tablenotes}
		\end{threeparttable}
	}   
\end{table}

\begin{table*}[ht]
	\caption{Sensitivity analyses of keypoint detection using different loss functions, backbone models, and keypoint masks.}
	\label{table:sensitivity_analysis}
	\centering
	\resizebox{0.96\textwidth}{!}{
		\begin{threeparttable}
			\begin{tabular}{@{}llllll@{}}
				\toprule
				& \textbf{ResNet50+FL+BM$^a$} & \textbf{ResNet50+FL+HM$^b$}  & \textbf{ResNet50+CEL$^c$+BM}  & \textbf{ResNeXt50$^d$+FL+BM} & \textbf{ResNet50+FPN+FL+BM}    \\ \midrule
				\begin{tabular}[c]{@{}l@{}}mAP\\ (95\% CI)\end{tabular} & \begin{tabular}[c]{@{}l@{}}\textbf{0.807}\\ (0.804--0.810)\end{tabular} &        \begin{tabular}[c]{@{}l@{}}0.804\\ (0.802--0.807)\end{tabular} &        \begin{tabular}[c]{@{}l@{}}0.794\\ (0.791--0.797)\end{tabular} & \begin{tabular}[c]{@{}l@{}}0.792\\ (0.788--0.797)\end{tabular} & \begin{tabular}[c]{@{}l@{}}0.799\\ (0.795--0.802)\end{tabular} \\ \\
				\begin{tabular}[c]{@{}l@{}}mAR\\ (95\% CI)\end{tabular}  & \begin{tabular}[c]{@{}l@{}}\textbf{0.870}\\ (0.867--0.872)\end{tabular} &        \begin{tabular}[c]{@{}l@{}}0.866\\ (0.863--0.868)\end{tabular} &        \begin{tabular}[c]{@{}l@{}}0.858\\ (0.856--0.861)\end{tabular} & \begin{tabular}[c]{@{}l@{}}0.858\\ (0.854--0.861)\end{tabular} & \begin{tabular}[c]{@{}l@{}}0.862\\ (0.859--0.864)\end{tabular} \\ \bottomrule
			\end{tabular}
			\begin{tablenotes}
				\item $^a$ ResNet50+FL+BM refers to the proposed model using ResNet50 as the feature backbone with focal loss (FL) and binary keypoint masks (BM). $^b$ HM denotes the heatmap keypoint mask. $^c$ CEL denotes the cross-entropy loss. $^d$ ResNeXt50 and FPN denote using the ResNeXt50 model and the Feature Pyramid Network as the feature backbone, respectively.
			\end{tablenotes}
		\end{threeparttable}
	}
\end{table*}

\begin{figure}[!h]
	\centering
	\begin{subfigure}[b]{0.95\columnwidth}
		\includegraphics[width=\linewidth]{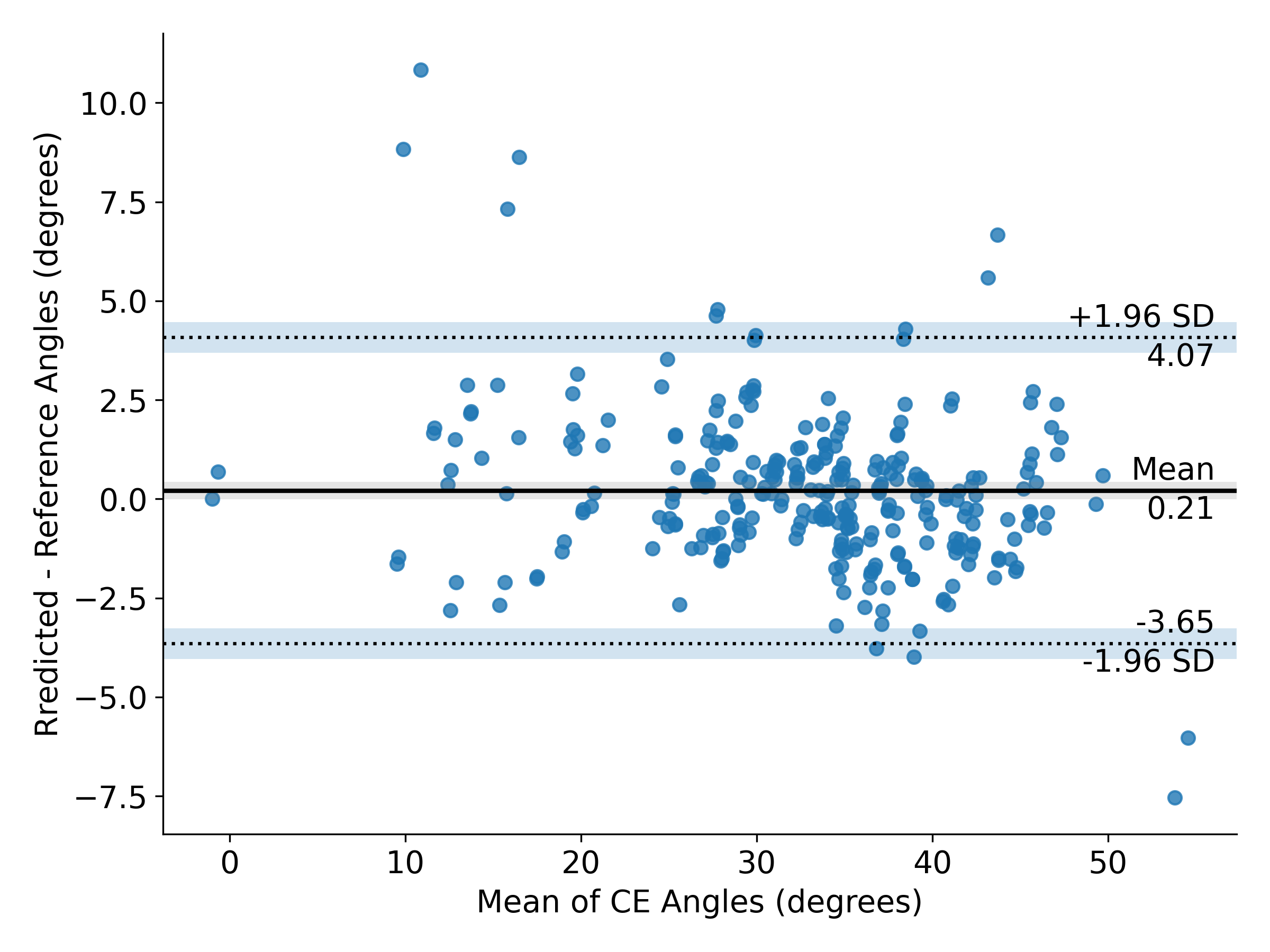}
		\caption{}
		\label{fig:3a}
	\end{subfigure}
	\begin{subfigure}[b]{0.95\columnwidth}
		\includegraphics[width=\linewidth]{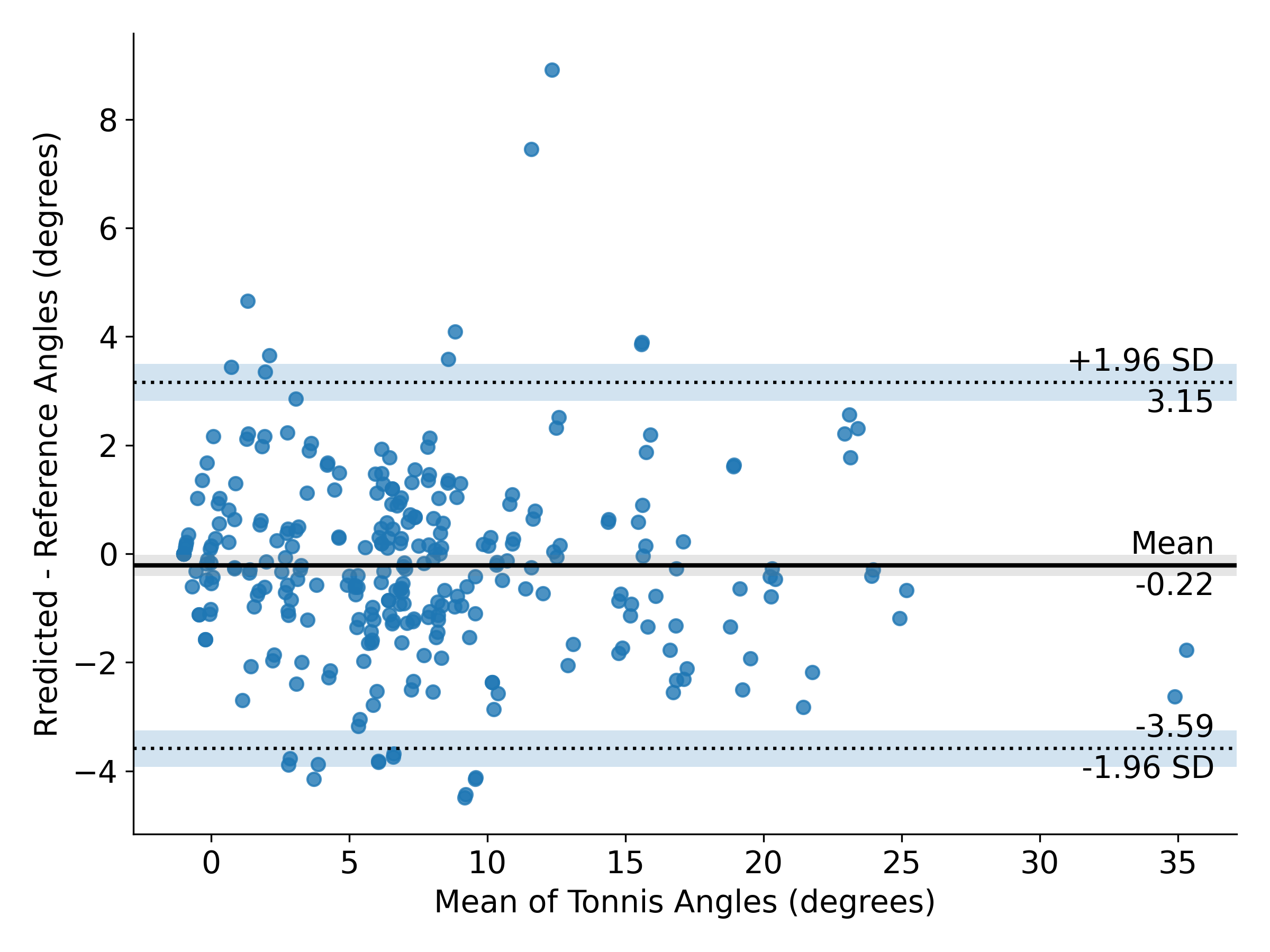}
		\caption{}
		\label{fig:3b}
	\end{subfigure}
	\begin{subfigure}[b]{0.95\columnwidth}
		\includegraphics[width=\textwidth]{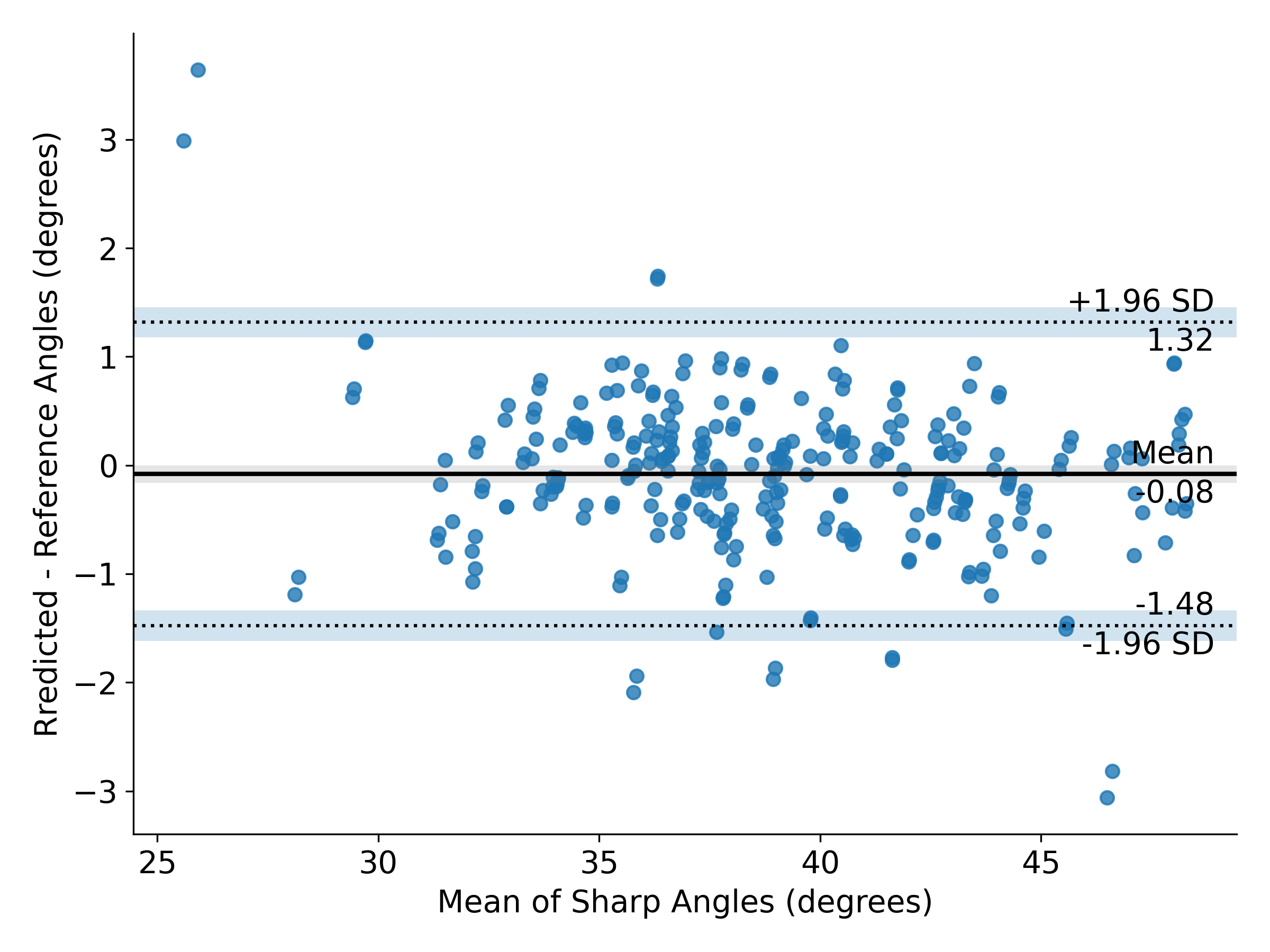}
		\caption{}
		\label{fig:3c}
	\end{subfigure}
	\caption{Bland-Altman analysis of the detected and reference measurements of the (a) Center-Edge (CE), (b) T\"onnis, and (c) Sharp angles in the Test set.}
	\label{fig:3}
\end{figure}

\begin{table*}[t]
	\caption{Comparison of intraclass correlation coefficients (ICC) of angle measurements\deleted{ between different groups in the Test set}.}
	\label{table:5}
	\centering
	\resizebox{0.90\textwidth}{!}{
		\begin{threeparttable}
			\begin{tabular}{@{}lllllll@{}}
				\toprule
				\textbf{Laterality} & \textbf{Angles} & \textbf{Our Model$^a$} & \textbf{Annotators$^b$} & \textbf{Orthopedists$^c$} & \textbf{Yang~\textit{et al.}\cite{yang2020feasibility}} & \textbf{Li~\textit{et al.}\cite{li2024deep}}\\ \midrule
				Right & CE & \begin{tabular}[c]{@{}l@{}}\textbf{0.965} (0.963--0.966)\end{tabular} & \begin{tabular}[c]{@{}l@{}}0.964 (0.946--0.983)\end{tabular} & \begin{tabular}[c]{@{}l@{}}0.875 (0.857--0.893)\end{tabular} & 0.86	&	0.908               \\
				& T\"onnis & \begin{tabular}[c]{@{}l@{}}0.950 (0.947--0.952)\end{tabular}  & \begin{tabular}[c]{@{}l@{}}\textbf{0.959} (0.938--0.980)\end{tabular} & \begin{tabular}[c]{@{}l@{}}0.917 (0.902--0.931)\end{tabular} & 0.83	&	0.790               \\
				& Sharp & \begin{tabular}[c]{@{}l@{}}\textbf{0.963} (0.961--0.965)\end{tabular}  & \begin{tabular}[c]{@{}l@{}}0.950 (0.921--0.979)\end{tabular} & \begin{tabular}[c]{@{}l@{}}0.919 (0.902--0.936)\end{tabular} & 0.93	&	0.943                 \\ \\
				Left  & CE & \begin{tabular}[c]{@{}l@{}}\textbf{0.949} (0.946--0.953)\end{tabular}  & \begin{tabular}[c]{@{}l@{}}0.910 (0.860--0.960)\end{tabular} & \begin{tabular}[c]{@{}l@{}}0.889 (0.877--0.902)\end{tabular} & 0.93	&	0.895                 \\
				& T\"onnis & \begin{tabular}[c]{@{}l@{}}\textbf{0.935} (0.932--0.937)\end{tabular}  & \begin{tabular}[c]{@{}l@{}}0.931 (0.895--0.967)\end{tabular} & \begin{tabular}[c]{@{}l@{}}0.876 (0.829--0.923)\end{tabular} & 0.86	&	0.757                 \\
				& Sharp & \begin{tabular}[c]{@{}l@{}}\textbf{0.969} (0.967--0.970)\end{tabular}  & \begin{tabular}[c]{@{}l@{}}0.924 (0.870--0.978)\end{tabular} & \begin{tabular}[c]{@{}l@{}}0.896 (0.887--0.906)\end{tabular} & 0.92	&	0.801                 \\ 
				\bottomrule
			\end{tabular}
			\begin{tablenotes}
				\item $^a$ \replaced{Our results in the Test set using models trained from}{Results of our model across} 10-fold cross-validation. \added{The data is presented in the form of the mean ICC (95\% confidence interval)}
				\item $^b$ Results of repeated measurements generated by the annotators. 
				\item $^c$ Results of eight orthopedists with over 6 years of clinical experience. 
			\end{tablenotes}
		\end{threeparttable}
	}
\end{table*}

\begin{figure}[ht]
	\centering
	\includegraphics[width=0.83\columnwidth]{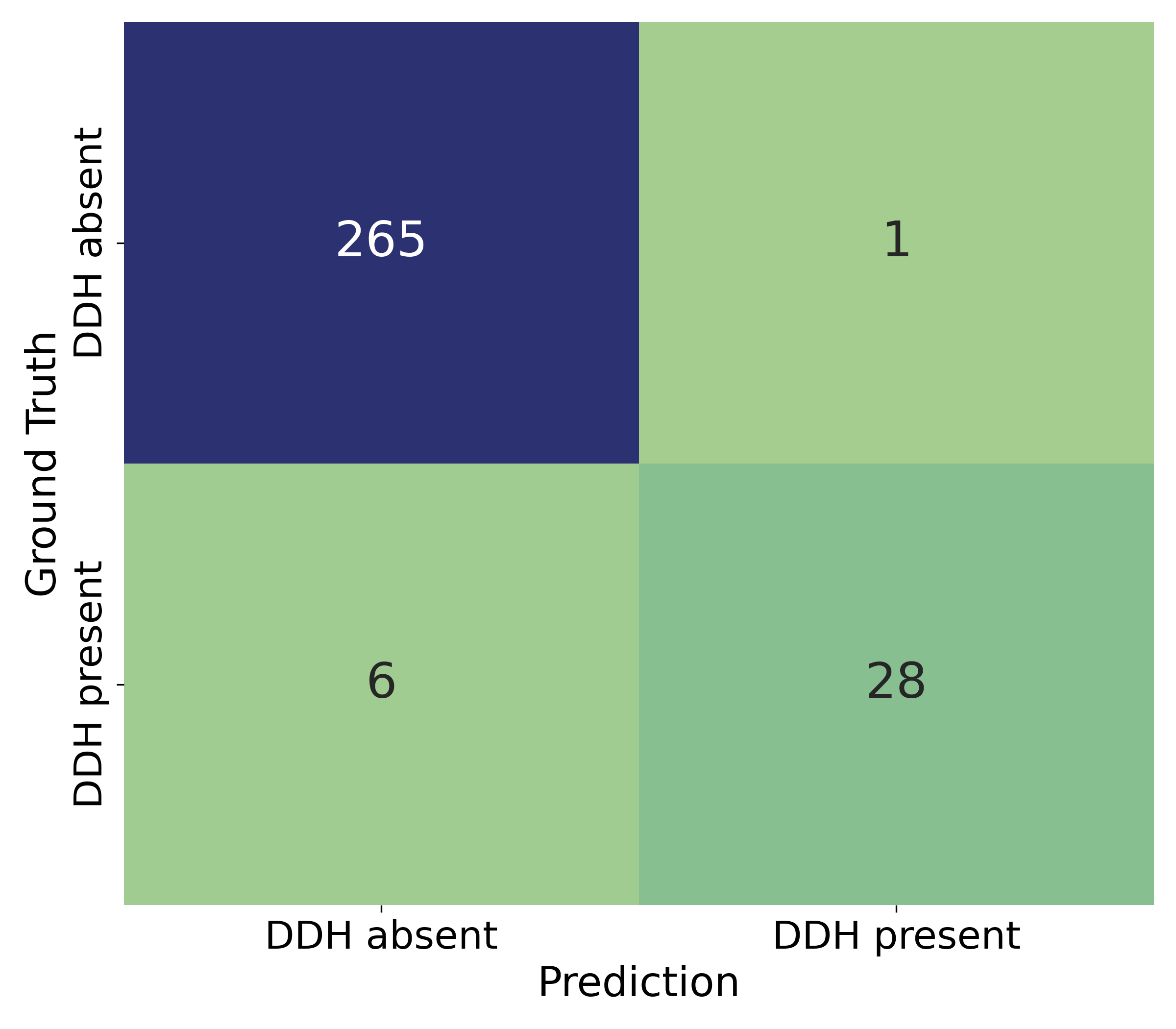}
	\caption{The confusion matrix of DDH diagnosis in the Test set using the proposed scoring system and the mean angle measurements across 10-fold cross-validation.}
	\label{fig:confusion_matrix}
\end{figure}

\section{Results}

A total of 1683 anteroposterior view pelvic radiographs (3366 hips) from 1683 patients (male: female = 623: 1060), with a mean age of 54.8 years (standard deviation: 18.5), were included in this study. The number of radiographs in the \replaced{Train-Val}{training} and \replaced{T}{t}est sets was 1533 and 150, respectively. The numbers of hips labeled as ``DDH absent" and ``DDH present" were 3024 and 342, respectively. Detailed data characteristics are summarized in Table~\ref{table:3}.

Using ResNet50 as the feature extraction backbone model, focal loss as the loss function, and binary keypoint masks as the training target (denoted as ResNet50+FL+BM), our keypoint detection model achieved an mAP of 0.807 (95\% CI: 0.804 to 0.810) and an mAR of 0.870 (95\% CI: 0.867 to 0.872), respectively. In comparison, models using alternative configurations such as cross-entropy loss, different backbone models (ResNeXt50 and ResNet50+FPN\footnote{ResNet50+FPN refers to the ResNet50 model with the feature pyramid network structure~\cite{lin2017feature}}), and heatmap keypoint masks consistently demonstrated inferior keypoint detection performance. As a result, the ResNet50+FL+BM model was used for all subsequent experiments. The detailed sensitivity analysis results for the keypoint detection models with different configurations are presented in Table~\ref{table:sensitivity_analysis}.

The Bland-Altman analysis for CE, T\"{o}nnis, and Sharp angles measured by our system and the ground truth measurements in the Test set are illustrated in Figure~\ref{fig:3}. The mean ICC for CE, T\"{o}nnis, and Sharp angles (for both sides) between our system and ground truth measurements were 0.957 (95\% CI: 0.952 to 0.962), 0.942 (95\% CI: 0.937 to 0.947), and 0.966 (95\% CI: 0.964 to 0.968), respectively. By comparison, the orthopedist group \added{with moderate clinical experience} achieved statistically significantly lower ICC in angle measurements ($p<0.001$), with 0.877 (95\% CI: 0.866 to 0.889), 0.894 (95\% CI: 0.865 to 0.922), and 0.906 (95\% CI: 0.894 to 0.917) for CE, T\"{o}nnis, and Sharp angles, respectively. Meanwhile, annotators' repeated measurements yielded mean ICCs of 0.944 (95\% CI: 0.913 to 0.974), 0.946 (95\% CI: 0.918 to 0.974), and 0.928 (95\% CI: 0.888 to 0.969) for CE, T\"{o}nnis, and Sharp angles, respectively, which were not significantly different from our results ($p=0.459$). Table~\ref{table:5} provides a detailed comparison of the angle measurement performance in the Test set, obtained by our model, annotators, \added{moderately-}experienced orthopedists, and state-of-the-art results~\cite{yang2020feasibility, li2024deep} for each side of the pelvis.

In terms of DDH diagnosis, when applying the scoring system to the three angles measured by our system (as described in Table~\ref{table:1}), the proposed diagnostic system achieved a mean F1 score of 0.863 (95\% CI: 0.851 to 0.876) in the Test set, which significantly outperformed that of the orthopedist group (0.777 [95\% CI: 0.737 to 0.817], $p=0.005$). When using the criteria for the three angles individually, the diagnostic performance was also significantly lower than our system ($p<0.001$), with the mean F1 scores for the CE, T\"{o}nnis, and Sharp angles of 0.790 (95\% CI: 0.783 to 0.797), 0.570 (95\% CI: 0.563 to 0.577), and 0.521 (95\% CI: 0.512 to 0.530), respectively. Additionally, the diagnostic F1 score can be further improved to 0.889 when using the model ensemble from the cross-validation. Figure~\ref{fig:confusion_matrix} illustrates the DDH diagnosis confusion matrix using our scoring system and the mean angle measurements obtained from models in the 10-fold cross-validation.

\section{Discussion} 

Radiography remains the primary imaging modality for early detection of developmental dysplasia of the hip (DDH). However, clinical DDH diagnosis relies heavily on manual evaluation of radiological landmark features, a process prone to subjectivity, inefficiency, and variability, especially in less experienced clinicians. In this study, we present a new deep learning-based system that automates DDH diagnosis from pelvic radiographs. This system integrates keypoint detection, radiological angle measurement, DDH diagnosis, and result visualization, offering a comprehensive and end-to-end solution. By combining the measurements of CE, T\"{o}nnis, and Sharp angles, our system achieved a significantly higher F1 score than moderately experienced clinicians' manual assessments, demonstrating its potential to enhance diagnostic accuracy and consistency.

Keypoint detection is an essential component of our system, as the accuracy of subsequent modules, including angle measurements and DDH diagnosis, highly depends on precise keypoint localization. We developed a modified Mask-RCNN architecture, replacing instance segmentation masks with ``one-hot” keypoint masks. To further refine keypoint detection, we introduced a parallel bounding box regression branch, which improved both mean average precision (mAP) and mean average recall (mAR), increasing mAP from 0.773 to 0.807 and mAR from 0.853 to 0.870. Moreover, using focal loss rather than cross-entropy loss allowed us to mitigate the impact of class imbalance in keypoint detection, leading to improved performance. Sensitivity analyses confirmed that our model (employing focal loss, ResNet50 for feature extraction,  binary keypoint masks, and bounding box regression) consistently outperformed other configurations (see Table~\ref{table:sensitivity_analysis}). While the original Mask-RCNN study by He \textit{et al.}~\cite{he2017mask} reported superior performance with more complex backbones like ResNet-FPN, we hypothesize that the relatively smaller data size in this study might limit the advantage of more sophisticated models. 

We utilized object keypoint similarity (OKS)-based mAP and mAR metrics to evaluate the performance of keypoint detection. OKS accounts for human variability in labeling the same keypoint, providing a perceptually meaningful assessment of the difference between detected and ground truth keypoints~\cite{ruggero2017benchmarking}. Our analysis of repeated annotations, which were used to estimate measurement variability among human experts, revealed substantial variation in labeling the medial aspect of the acetabulum (keypoint D, Figure~\ref{fig:1}), with variability levels two to three times higher than those for the femoral head center (keypoint B, Figure~\ref{fig:1}). This disparity suggests that clinical measurements reliant on the medial aspect of the acetabulum, such as the T\"{o}nnis angle, may not as reliable as those based on the femoral head center, such as the CE angle---a finding that aligns with the clinical preference for CE angle in DDH diagnosis.

The ICC of angle measurements generated by our model was comparable to that of repeated measurements from expert annotators, indicating that our model achieves accuracy on par with highly experienced orthopedic surgeons (with over 15 years of clinical experience). Furthermore, our model demonstrated lower variance in those angle measurements than human annotators, as reflected in the narrower confidence intervals of ICC values in Table~\ref{table:5}. This consistency highlights the robustness of our system in providing reliable measurements, a critical factor in clinical decision-making. Additionally, the ICC values for our system were statistically significantly higher than those obtained by moderately experienced orthopedists and prior state-of-the-art models, underscoring the system's superior performance.

\begin{figure}
	\centering
	\includegraphics[width=0.95\columnwidth]{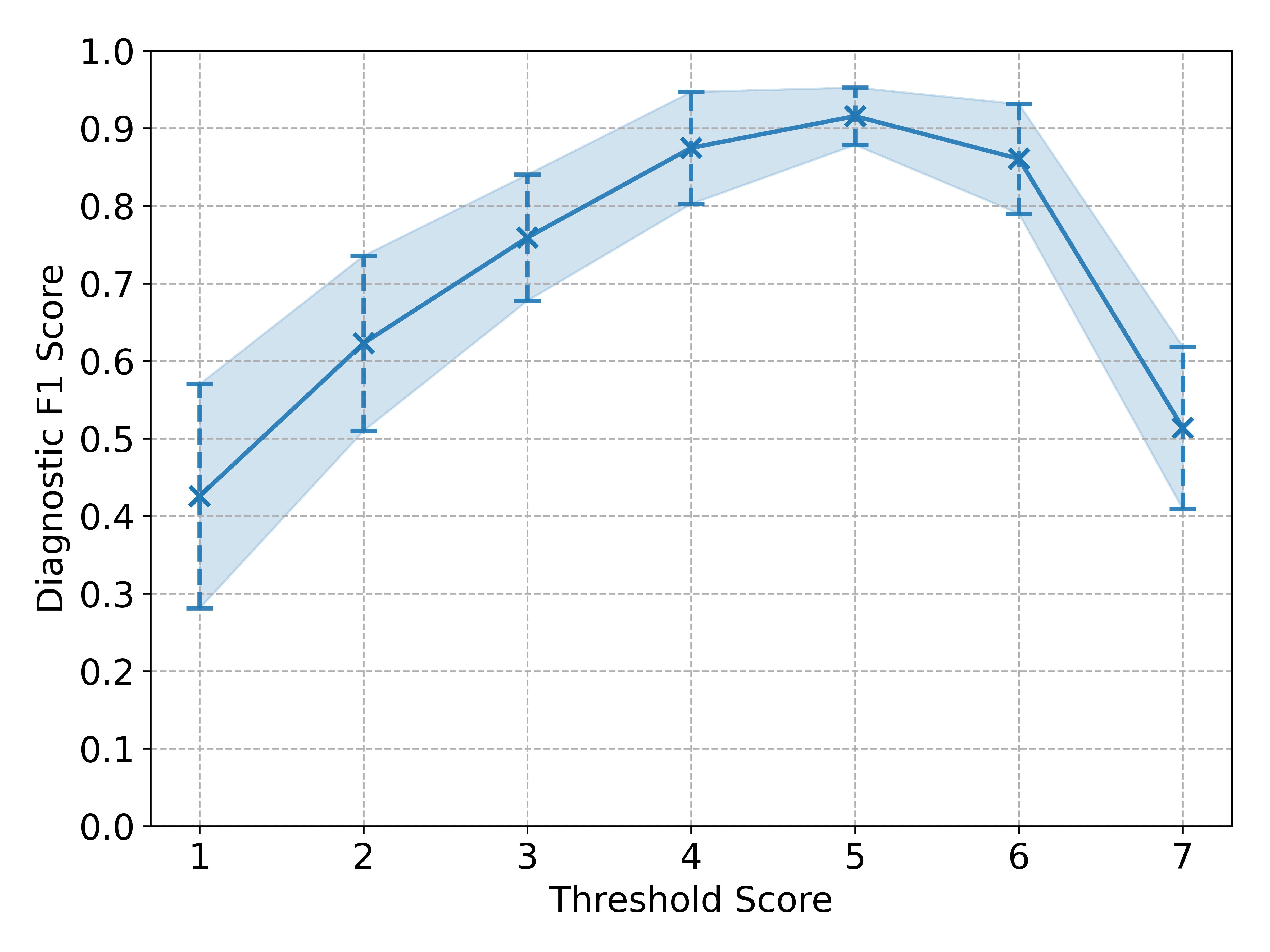}
	\caption{Relationship between the diagnostic threshold of the scoring system \added{(x-axis)} and the diagnostic \replaced{F1 score (y-axis)}{performance (F1 score)}. \added{The solid lines connect the mean F1 score using different threshold values over the 10-fold cross-validation grid search. The error bar and shaded area represent the range of plus-minus 1-time standard deviation across the 10-fold cross-validation.}}
	\label{fig:5}
\end{figure}

To quantitatively integrate information from the CE, T\"onnis, and Sharp angles, we developed a data-driven scoring system for comprehensive DDH diagnosis. This system assigns different weights based on diagnostic criteria for each angle, with the final diagnosis determined by the cumulative score. We conducted \added{a 10-fold cross-validation} grid search\deleted{es} \added{in the Train-Val set} to optimize the scoring system parameters, including the individual scores assigned to each angle and the total diagnostic threshold\deleted{, to maximize diagnostic performance (F1 score) in the training set}. \added{This 10-fold CV search reduces the risk of overfitting and provides a more generalized evaluation of the scoring system parameters.} For example, for the CE angle, a score of 3 yielded the highest performance (mean F1 score = 0.913), compared to scores of 1, 2, and 4, which achieved F1 scores of 0.832, 0.878, and 0.886, respectively. Therefore, we selected a score of 3 for the CE angle in our DDH diagnostic system. \deleted{Moreover, }Figure~\ref{fig:5} illustrates the selection process for the total threshold score, where a score of 5 provided the optimal outcome \added{(i.e., the highest mean F1 score and the lowest diagnostic variance over the 10-fold CV)}. Importantly, all parameters in the proposed system were derived from data-driven diagnostics using a reasonably large dataset rather than relying on handcrafted rules. \added{Moreover, unlike previous deep learning models that function as “black boxes” with limited explainability, our system transparently maps input measurements to diagnostic outcomes by explicitly defining how each radiological angle contributes to the final decision.} As a result, the scoring system enhances both the interpretability and generalizability of DDH diagnosis, providing a clear, self-contained explanation \replaced{to clinicians for a better understanding of the reasoning behind each diagnosis}{for each decision}. 

Furthermore, the proposed scoring system prioritizes abnormal CE angles over T\"onnis or Sharp angles (Table~\ref{table:1}). This behavior is consistent with findings in the literature~\cite{pereira2014recognition, welton2023radiographic} as well as clinical practice, which can further validate the credibility and explainability of our system’s diagnoses. In terms of diagnostic performance, our system handled the imbalanced Test set effectively, with a specificity of 0.996 and a sensitivity of 0.824 (see Figure~\ref{fig:confusion_matrix}). It also significantly outperformed a cohort of moderately experienced orthopedists (Mann–Whitney U test $p=0.005$). In addition, the mean diagnosis F1 score of our system (0.863) considerably exceeded the results reported by previous work~\cite{li2019auxiliary}, where the diagnosis was based solely on the Sharp angle (F1 score = 0.312). This highlights the importance of integrating multiple angles to improve diagnostic accuracy in DDH.

\added{With automated and reliable angle measurements and DDH diagnosis, the proposed system could serve as a valuable clinical decision-support tool, particularly for less-to-moderately experienced clinicians and complex cases. By providing consistent assessments, our system may also facilitate earlier detection and timely intervention, potentially preventing disease progression and reducing the need for invasive treatments. Furthermore, in remote or underserved regions with limited access to orthopedic specialists, using such AI-driven systems could enable timely online consultations and second-opinion assessments, promoting more equitable healthcare delivery. Future studies are needed to thoroughly evaluate its application in real-world clinical settings and assess its impact on patient outcomes and healthcare workflows.}

Despite these promising results, there are limitations to consider. First, the scoring system for DDH diagnosis was developed and evaluated using data from a single center. Although the performance was tested on a set of unseen data, the single source data may introduce biases related to the specific clinical practices of that institution. \deleted{Second, the diagnostic labels were based on subjective evaluations and manual measurements by annotators, potentially introducing inter-reader variability that could affect performance assessments.} \replaced{Additionally}{Lastly}, the relatively small data\deleted{set} size may have limited the ability to \replaced{explore}{train} more sophisticated deep learning models, such as more complex feature extraction backbones in keypoint detection. As such, future work will focus on collecting additional and external data from multiple sources \added{with ground truth labels generated by different clinicians} to validate and enhance the generalizability of our proposed system. Moreover, different clinical applications of our system, such as the interactive or cooperative diagnosis, would also warrant future investigation. \added{Lastly, while our scoring system effectively integrates multiple radiological angles, its performance may be influenced by varying or evolving threshold definitions, particularly for mild and borderline cases. To that point, future work should explore adaptive refinements to the scoring system and validate its robustness across different clinical guidelines.}

\section{Conclusion}
In this study, we presented a fully automated end-to-end system for comprehensive DDH diagnosis from pelvic radiographs based on deep learning keypoint detection and a new data-driven scoring system. The proposed approach demonstrated state-of-the-art performance on different tasks and can be used to provide reliable and explainable support for DDH diagnosis.

\vspace*{0.5 cm}


\vspace*{ 1 cm}

\bibliographystyle{ieeetr}
\bibliography{references}
%

%
%
%
%
%
%
%
\end{document}